\documentclass[namedreferences]{solarphys_arxiv}
\usepackage[optionalrh,solaromanenum,natbib]{arxiv-addons}
\usepackage[pdfborder={0 0 0 },urlcolor=blue]{hyperref}
\ifx \doiurl \undefined \def \doiurl#1{\href{http://dx.doi.org/#1}{\url{#1}}}\fi
\ifx \adsurl \undefined \def \adsurl#1{\href{http://adsabs.harvard.edu/abs/#1}{\url{#1}}}\fi
\usepackage{graphicx}
\usepackage{url}
 
\newcommand{\arcsec}{\hbox{$^{\prime\prime}$}}
\begin{document}
\begin{article}
\begin{opening}
\title{Two-Dimensional Helioseismic Power, Phase, and Coherence Spectra of {\it Solar Dynamics Observatory} Photospheric and Chromospheric Observables
}

\author{Rachel Howe$^{1}$\sep
        Kiran Jain$^{2}$ \sep
        Richard~S.~Bogart$^{3}$\sep 
        Deborah~A.~Haber$^{4}$\sep
        Charles~S.~Baldner$^{3}$ 
}
\runningauthor{R. Howe {\it et al.}} 

\runningtitle{Local Spatio--Temporal Power and Phase for HMI and AIA UV Observables}        

\institute{$^{1}$
              School of Physics and Astronomy,
              University of Birmingham,
              Edgbaston, Birmingham B15 2TT, UK
              email: \url{rhowe@nso.edu}             \\
                     $^{2}$National Solar Observatory, 950 N. Cherry Avenue, Tucson AZ 85719, USA\\
$^{3}$Stanford University,
Stanford, CA 94305-4085,  USA \\
$^{4}$JILA, University of Colorado, Boulder, CO, USA
}



\begin{abstract}
While the {\it Helioseismic and Magnetic Imager} (HMI) onboard the {\it Solar Dynamics Observatory} (SDO) provides Doppler velocity [$V$], continuum intensity [$I_C$], and line-depth [$Ld$] observations, each of which is sensitive to the five-minute acoustic spectrum, the {\it Atmospheric Imaging Array} (AIA) also observes at wavelengths -- specifically the 1600 and 1700 Angstrom bands -- that are partly formed in the upper photosphere and have good sensitivity to acoustic modes. In this article we consider the characteristics of the spatio--temporal Fourier spectra in AIA and HMI observables for a 15-degree region around NOAA Active Region 11072. We map the spatio--temporal-power distribution for the different observables and the HMI Line Core [$I_L$], or Continuum minus Line Depth, and the phase and coherence functions for selected observable pairs, as a function of position and frequency. Five-minute oscillation power in all observables is suppressed in the sunspot and also in plage areas. Above the acoustic cut-off frequency, the behaviour is more complicated: power in HMI $I_C$ is still suppressed in the presence of surface magnetic fields, while power in HMI $I_L$ and the AIA bands is suppressed in areas of surface field but enhanced in an extended area around the active region, and power in HMI~$V$ is enhanced in a narrow zone around strong-field concentrations and suppressed in a wider surrounding area. The relative phase of the observables, and their cross-coherence functions, are also altered around the active region. These effects may help us to understand the interaction of waves and magnetic fields in the different layers of the photosphere, and will need to be taken into account in multi-wavelength local helioseismic analysis of active regions.
%
%
%

\end{abstract}

\end{opening}

\section{Introduction}

\label{intro}

Helioseismology relies on the observation of waves, as they affect light from the
outer layers of the Sun, to infer the structure and dynamics of the otherwise
invisible deeper layers. By studying the oscillations in multiple wavelengths of light -- and hence at different heights in the atmosphere -- simultaneously, 
it is possible to probe the layers of the atmosphere. Such investigations,
aiming both to better understand the behaviour of the waves and their
interaction with magnetic fields, and to improve the inference of subsurface
properties, have a long history.

The launch in February 2010 of the {\it Solar Dynamics Observatory} (SDO), carrying both the
{\it Helioseismic and Magnetic Imager} (HMI) and the {\it Atmospheric Imaging Assembly} (AIA), provides new opportunities for cross-spectral helioseismic analysis, with full-disc, high-cadence images at many UV and EUV wavelengths from AIA as well as photospheric Doppler-velocity, continuum, and magnetic data from HMI. As discussed by \citet{2011JPhCS.271a2058H}, 
the AIA {1600\ \AA} and {1700\ \AA} near-ultraviolet bands show a clear signature of the
five-minute acoustic spectrum that  -- at least in the Sun-as-a-star case -- is much less contaminated by 
granulation ``noise'' than the continuum intensity in HMI's visible 6173~{\AA} line.

Both 
the effect of surface magnetic activity on helioseismic waves
in local areas and
the phase and coherence relationships between
the velocity and the intensity of radiation at different wavelengths 
have been studied for nearly two decades. 

\begin{figure}
\includegraphics[width=11.0cm]{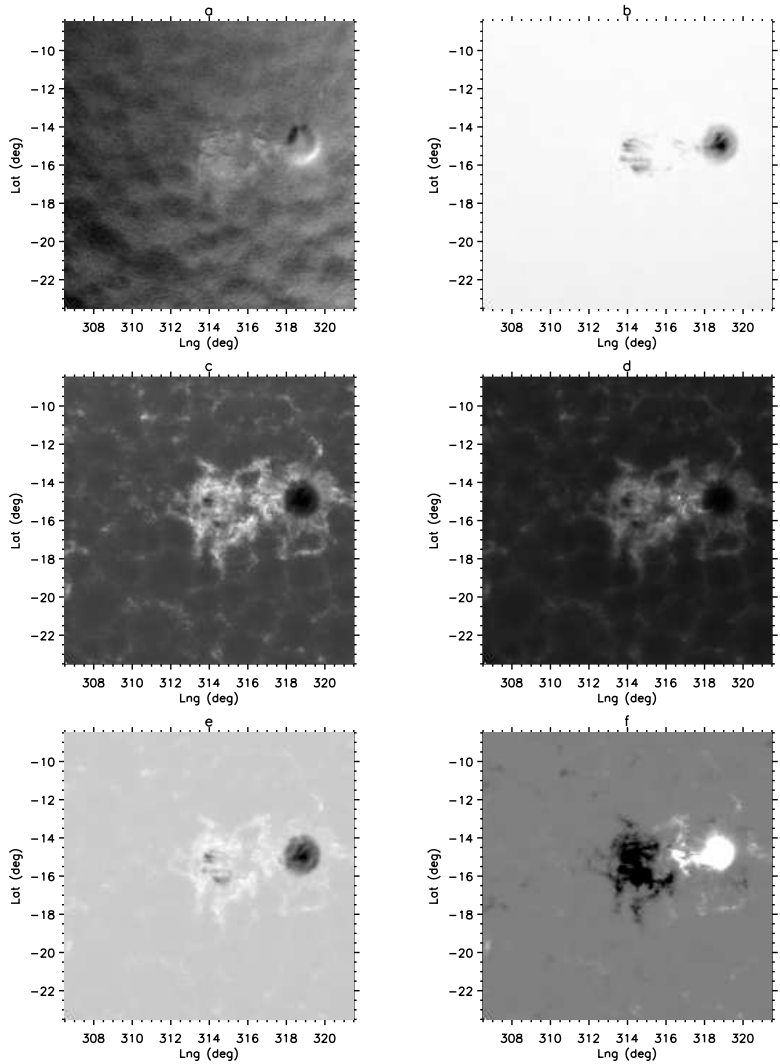}
\caption{ Spatial variation of each observable averaged over 23 May 2010: 
HMI~$V$ (a), HMI $I_C$ (b), AIA {1600 \AA} (c), AIA {1700 \AA} (d), HMI $I_L$ (e),
and HMI longitudinal magnetic-field strength (f).}
\label{fig:1}
\end{figure}

\begin{figure}
\includegraphics[width=11.0cm]{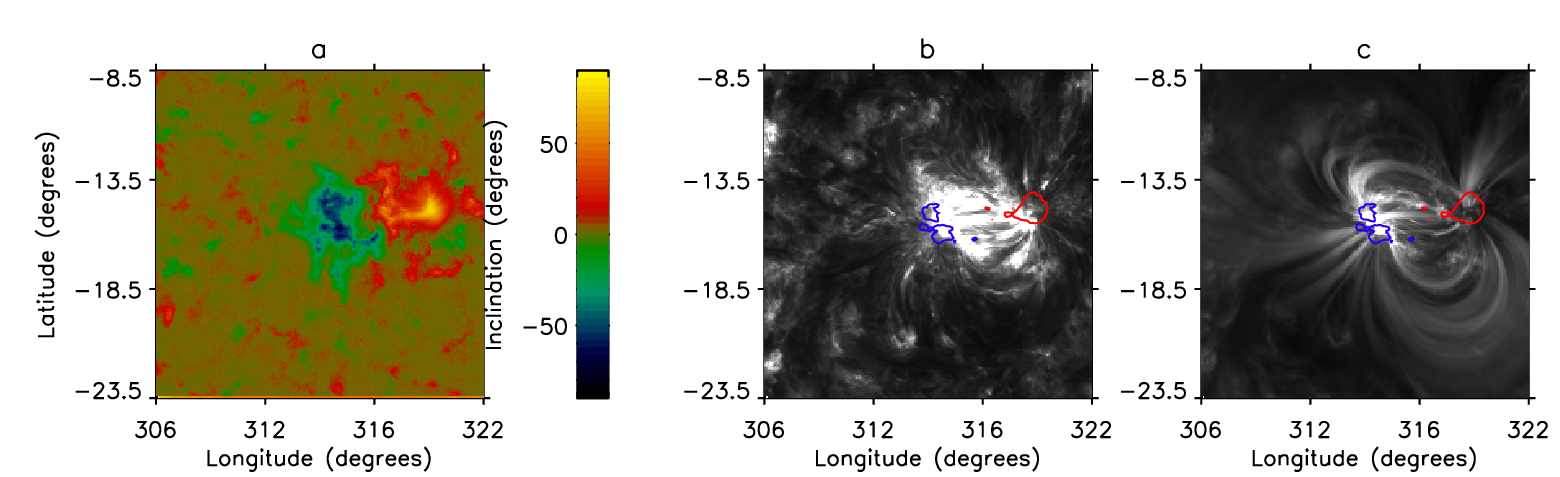}
\caption{The mean magnetic inclination (a), from HMI vector magnetograms, courtesy K. Hayashi, an AIA {304\ \AA} image (b), and 
an AIA {171\ \AA} image (c), for the area of interest on 23 May 2010. The 
contours overlaid on panels a and b show the {500\ G} level of the mean magnetic field strength from HMI line-of-sight magnetograms.}
\label{fig:1a}
\end{figure}

\citet{1992ApJ...394L..65B} reported finding small areas associated
with active regions that produced a disproportionate amount 
of acoustic power in the 5.5\,--\,7.5 mHz frequency band in ground-based Doppler observations using the Fe~{\sc i}~{5576\ \AA} line, while \citet{1992ApJ...392..739B}
made the first observations of defined acoustic haloes around active regions in the same frequency band using Ca~K intensity. However, atmospheric seeing can
have confusing effects on ground-based intensity observations \citep{2001ESASP.464..219H}, particularly in sunspot umbrae. The picture became clearer with the launch of the {\it Michelson Doppler Imager} \citep[MDI:][]{1995SoPh..162..129S} onboard the {\it Solar and Heliospheric Observatory} (SOHO), which allowed full-disc observations from space in 
both intensity and velocity using the {6768\ \AA} Ni line.
\citet{2002A&A...387.1092J} studied the power distribution in the
line-depth, Doppler-velocity, and continuum-intensity observables from
MDI. In both sunspot and plage regions they found that the power in areas of strong
magnetic field  was suppressed in all observables, while for
velocity and line depth (but not for continuum intensity) there was a ``halo'' of enhanced power surrounding the magnetic-field concentrations, which they concluded 
was acoustic in origin.
The {\it Transition Region and Coronal Explorer} (TRACE)  
made possible high-resolution, space-based observations in the 1600 and {1700\ \AA} bands.
\citet{2001ApJ...554..424J} used data from another SOHO instrument, {\it Solar Ultraviolet Measurements of Emitted Radiation} (SUMER), TRACE, and MDI to investigate
the oscillations in the chromosphere. This work established 
that the chromospheric modes were
primarily the same $p$-modes seen in the photosphere.
\citet{2001A&A...379.1052K} used TRACE to study oscillations in the {1600\ \AA} UV band
in the quiet Sun and found enhanced power at three-minute periods around
patches of network field. 
\citet{2003A&A...401..685M} also looked at acoustic power around an active region in 
short (two\,--\,four hour) stretches of TRACE data and did not find a high-frequency acoustic halo in the
1600 or {1700\ \AA} bands. 

The {\it Hinode/Solar Optical Telescope} has also been used for multi-wavelength
seismology using the Ca~H line and the G band, for example by \citet{2007PASJ...59S.637S}, who considered power, phase, and coherence spectra but not spatial maps, and by
\citet{2007PASJ...59S.631N} who looked at the power distribution in and around a sunspot. In the latter article there was no high-frequency acoustic halo detected, but there was a narrow band of excess power surrounding the sunspot at all wavelengths in the G band, and pronounced suppression of power at both wavelengths in the penumbra and plage, while the umbra showed excess power in the H line, especially above {4.5\ mHz}. 

\citet{2011SoPh..268..349S} have examined the power distribution 
for velocity observations from MDI and found that the excess high-frequency
power corresponds to regions with the magnetic-field inclination 
(as deduced from potential-field source-surface extrapolation) in the 40\,--\,60 degree range. However, the
exact origin of these high-frequency haloes remains unclear.

The relative phase of the Doppler velocity and the intensity in 
various wavelength bands has been studied for more than two decades in both resolved and unresolved observations. Early (ground-based) spatially resolved phase measurements were made, for example, by \citet{1990A&A...236..509D, 1992A&A...266..560D}; in the latter article they used three different wavelengths including the
Na D line, formed in the lower chromosphere. These measurements revealed a complicated pattern of phase relationships that was not easily explained by simple models.
\citet{2007A&A...471..961M} made ground-based observations from the South Pole
using the MOTH instrument \citep{2004SoPh..220..317F}, taking high-cadence
images in the sodium and potassium D lines. The sodium line shows clear 
evidence of a high-frequency halo around an active region.


\begin{figure}
\includegraphics[width=11.0cm]{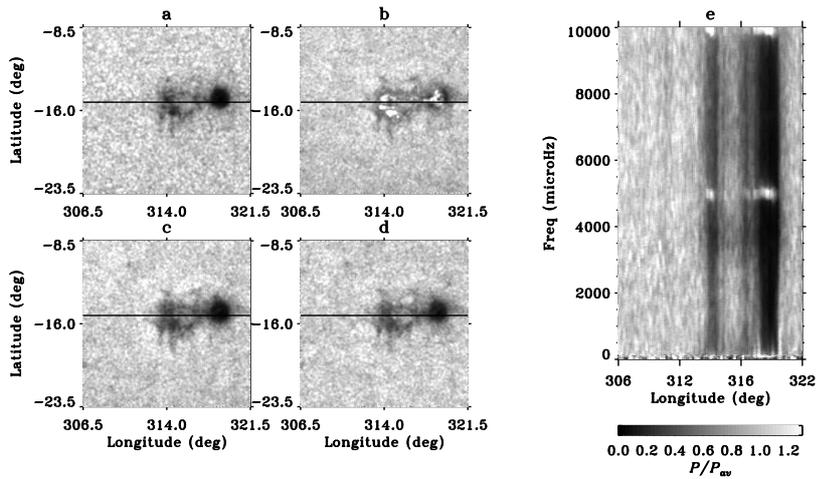}
\caption{Slices through the smoothed relative-power maps for HMI $I_C$, 
at 3\ mHz (a), 5\ mHz (b), 7\ mHz (c), 9\ mHz (d), and in longitude and temporal frequency along the
horizontal line shown in panels a\,--\,d (e). The grey-scale bar applies to all of the panels.}
\label{fig:2}
\end{figure}
\begin{figure}
\includegraphics[width=11.0cm]{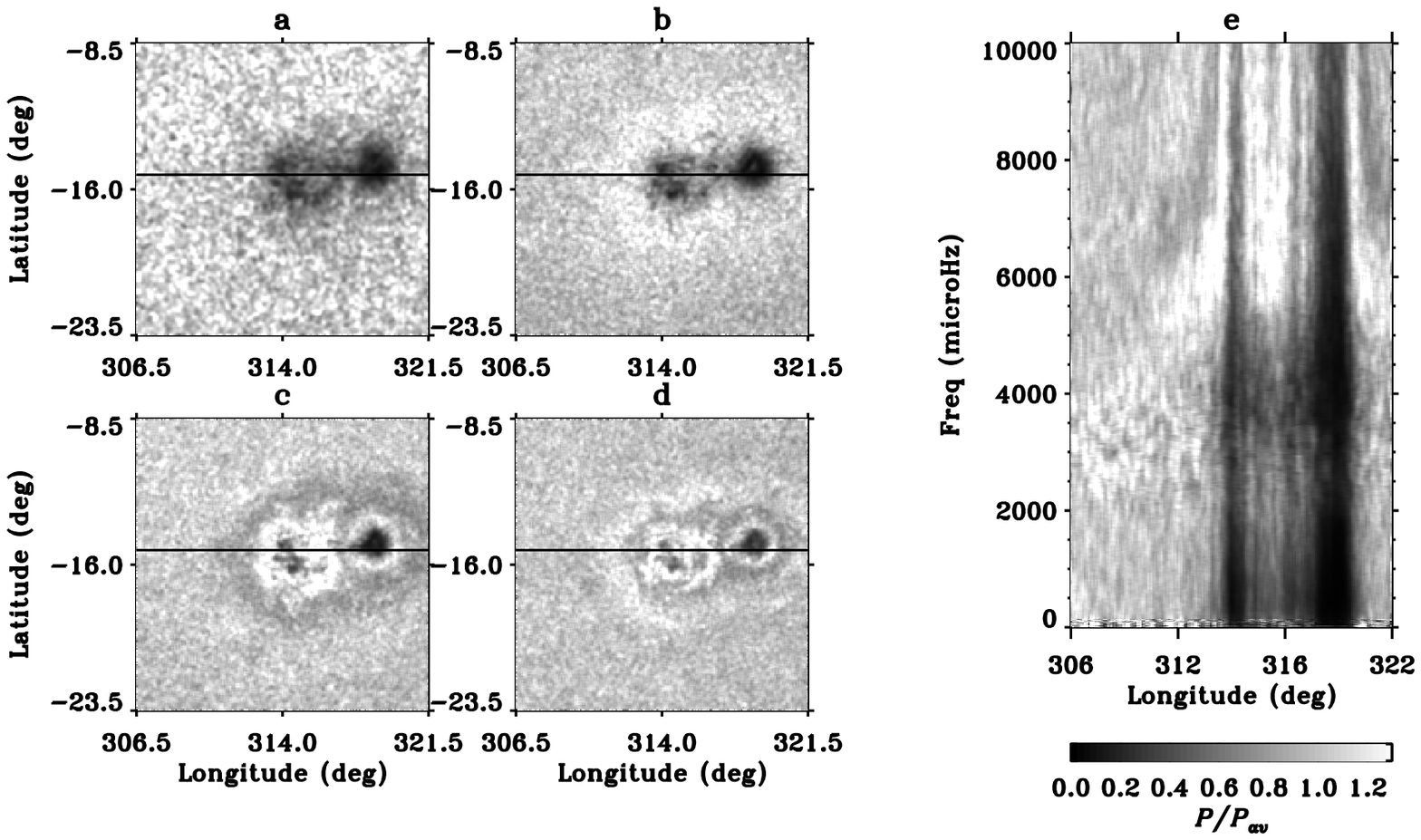}
\caption{Slices through the smoothed relative-power maps for HMI~$V$ at 3\, mHz (a), 5\ mHz (b), 7\ mHz (c), 9\ mHz (d), and in longitude and temporal frequency along the
horizontal line shown in panels a\,--\,d (e). The grey-scale bar applies to all of the panels.}

\label{fig:3}
\end{figure}

\begin{figure}
\includegraphics[width=11.0cm]{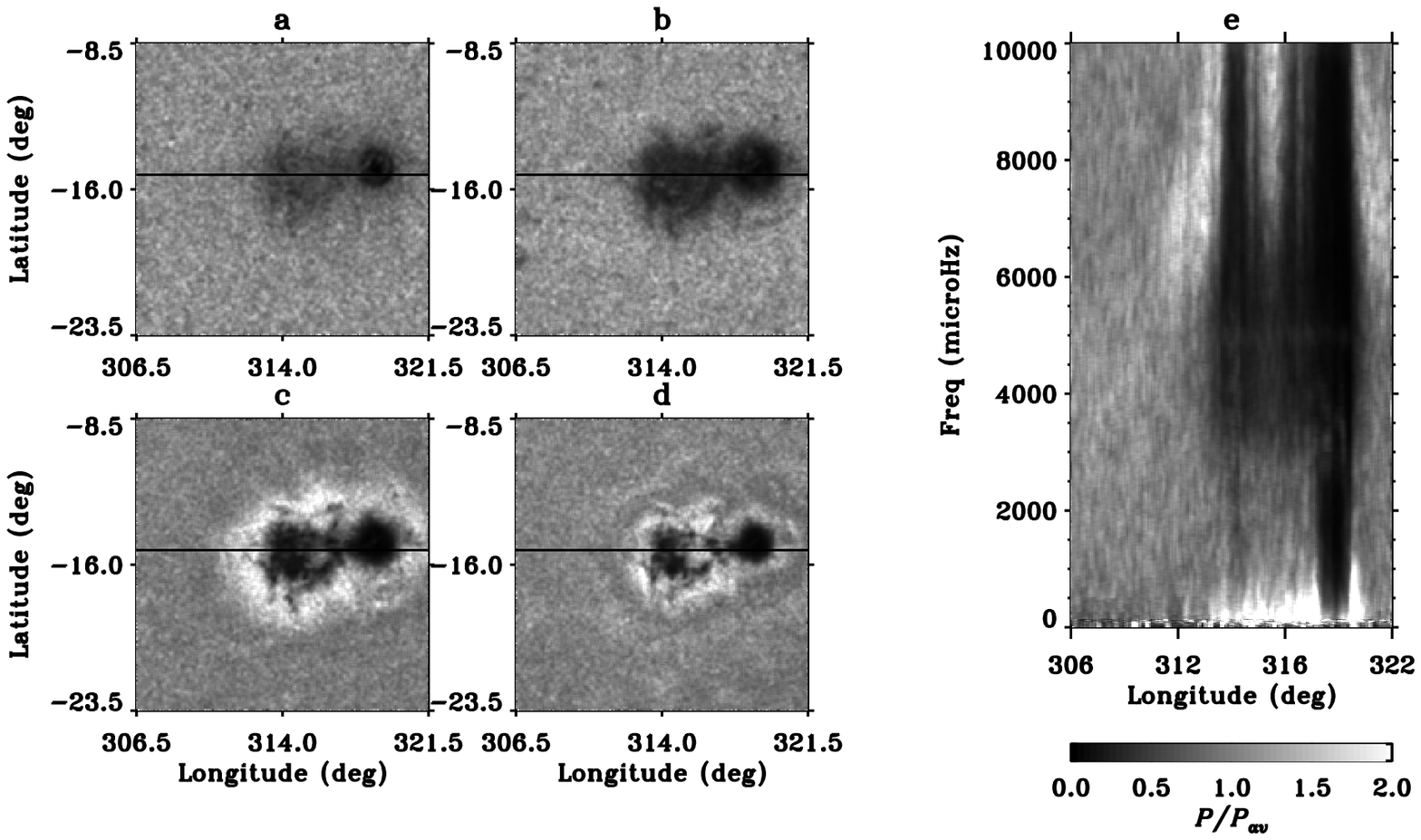}
\caption{Slices through the smoothed relative-power maps for HMI $I_L$,
at 3\ mHz (a), 5\ mHz (b), 7\ mHz (c), 9\ mHz (d), and in longitude and temporal frequency along the
horizontal line shown in panels a\,--\,d (e). The grey-scale bar applies to all of the panels.}
\label{fig:4}
\end{figure}

\begin{figure}
\includegraphics[width=11.0cm]{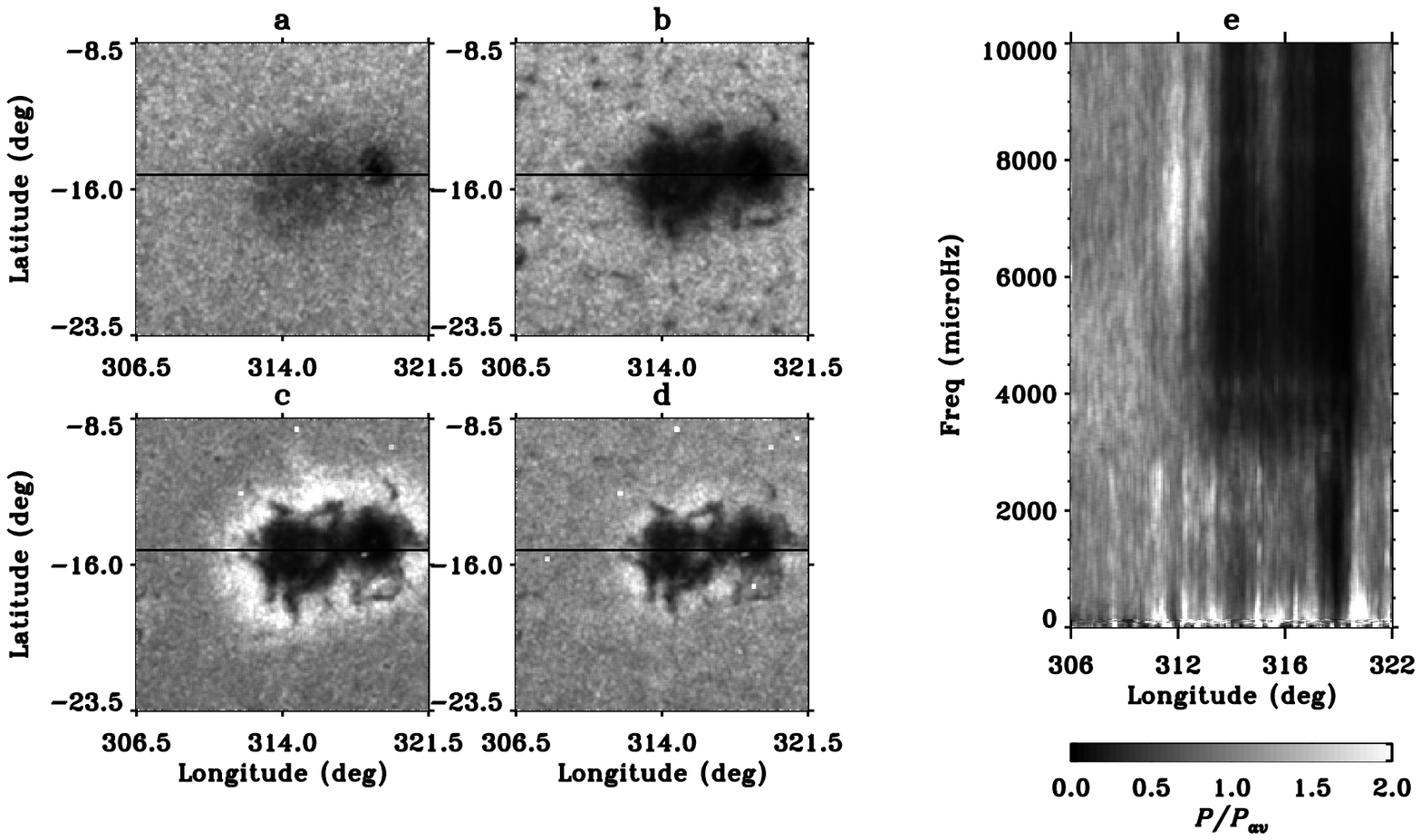}
\caption{Slices through the smoothed relative-power maps for {AIA 1700\ \AA},
at 3\ mHz (a), 5\ mHz (b), 7\ mHz (c), 9\ mHz (d), and in longitude and temporal frequency along the
horizontal line shown in panels a\,--\,d (e). The grey-scale bar applies to all of the panels. }
\label{fig:5}
\end{figure}

\begin{figure}
\includegraphics[width=11.0cm]{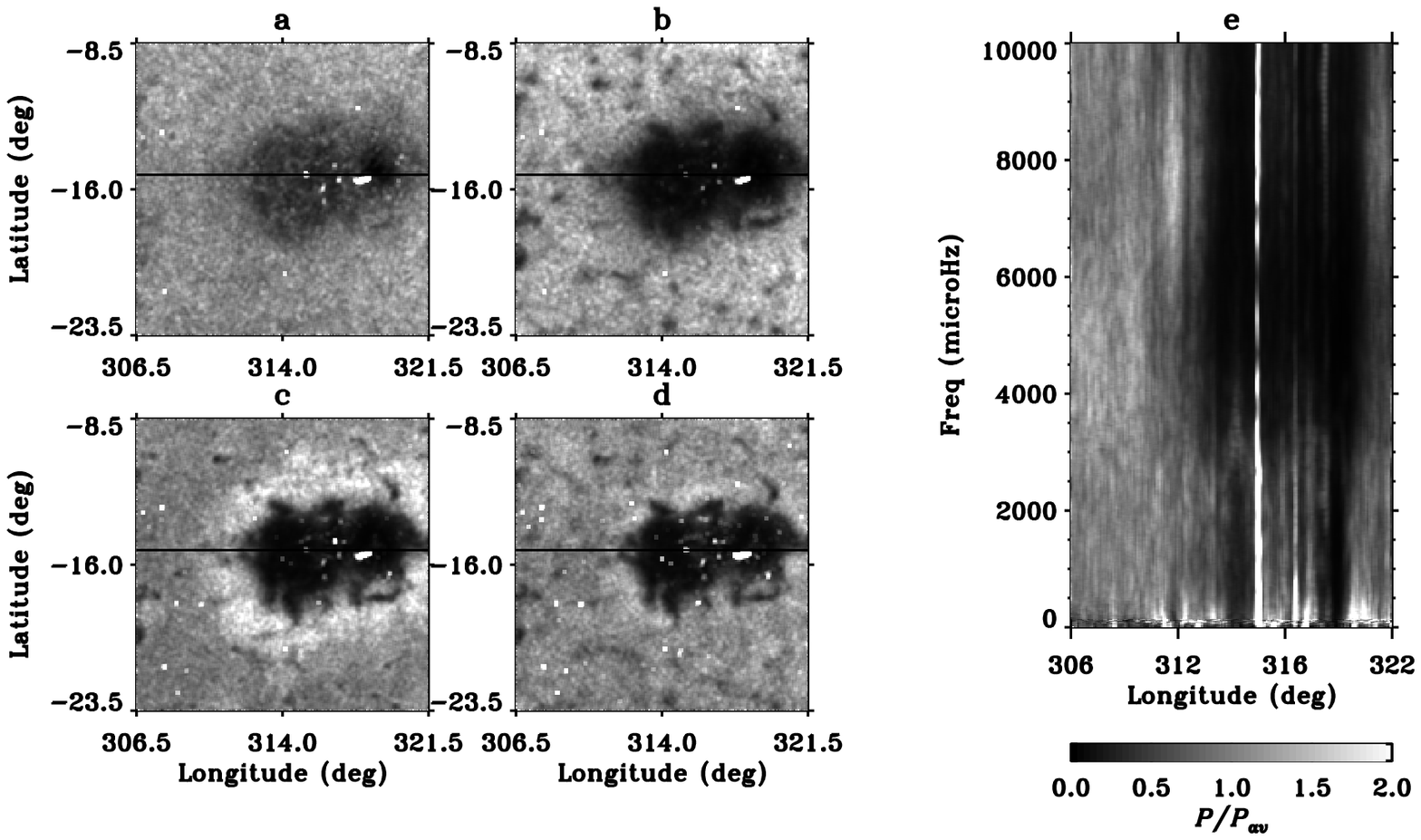}
\caption{Slices through the smoothed relative-power maps for {AIA 1600\ \AA}, at 3\ mHz (a), 5\ mHz (b), 7\ mHz (c), 9\ mHz (d), and in longitude and temporal frequency along the
horizontal line shown in panels a\,--\,d (e). The grey-scale bar applies to all of the panels.}
\label{fig:6}
\end{figure}


In this article we will examine the behaviour and relationships of the different observables in a small region of the solar surface containing an active region as observed by HMI and AIA, and show how the power and phase of the oscillations is affected by the
presence of local magnetic fields. For this purpose we will show observations of NOAA Active Region 11072 on 23 May 2010. We have also examined the 
data for the same region over the two preceding and following days; the results are very similar and for reasons of space are not shown in this article. 

In Section~\ref{sec:2} we will describe our data and analysis and define the phase and coherence spectra. In Section~\ref{sec:3} we show the region in each of the observables; in Section~\ref{sec:4} we present the power maps for each
observable, and in Section~\ref{sec:5} we show the phase and coherence spectra. In Section~\ref{sec:6} we discuss our findings.

\section{Data and Analysis}
\label{sec:2}
\subsection{Helioseismic Observables from the {\it Solar Dynamics Observatory}}
\label{sec:2.1}

The HMI observables are produced from observations of the 
Fe~{\sc i} line at {6173~\AA} by combining ``filtergrams'' taken with different configurations of the instrument's filters. 
For each 45-second interval there is a line-of-sight magnetogram, 
a Doppler-velocity image [HMI~$V$], a continuum intensity image [HMI $I_C$],
a Line Depth image [HMI Ld], and also a Line Width image, which is not considered here. For this investigation
we consider the HMI $I_C$ and HMI~$V$ observables and also a Line Core intensity observable [HMI $I_L$] formed by taking the difference between HMI $I_C$ and HMI Ld. The pixel size of the $4096\times4096$ pixel images is approximately {0.5\arcsec}, with an optical resolution of {1\arcsec}.

\citet{2006SoPh..239...69N} found that the height of formation of the 
HMI {6173\ \AA} line spans the range from {20\ km} at the wings to {270\ km} at the core, while 
\citet{2011SoPh..271...27F} have calculated that the height of formation
of the HMI Doppler velocity is approximately {100\ km}, slightly lower than that of the Ni~{6768\ \AA} line used by MDI. However, the finite resolution of the
instruments raises the effective height of formation by about {50\ km} at disc centre in both cases.
We can therefore say that the HMI $I_C$ observable is formed considerably lower in the atmosphere than HMI $I_L$, with HMI~$V$ at an intermediate depth. All of these heights apply to quiet Sun, and may well be altered in active regions.

The AIA {1600\ \AA} images for the day analysed have a cadence of 24 seconds, while the cadence for {1700\ \AA} 
was 48 seconds. (More recent data have used a consistent 
24-second cadence for both, but the cadences were adjusted several times 
during the early months of AIA observations.) These images are also taken at $4096\times4096$ pixels, but because AIA allows more room around the limb of the solar image, the pixel size is approximately {0.6\arcsec}. The {1700\ \AA} band is believed to be formed in the upper photosphere; \citet{2001ApJ...554..424J} put it at 300\,--\,550 km, around the temperature minimum between the
upper photosphere and lower chromosphere. The {1600\ \AA} band includes the C~{\sc iv} line and samples both the 
transition region at the top of the chromosphere and the upper 
photosphere. The calculations of \citet{2005ApJ...625..556F} for the
equivalent TRACE bandpasses give mean formation heights of {360\ km} and 
{480\ km} for the 1700 and {1600\ \AA} bands, with FWHM of 385 and {185\ km} respectively; these authors also note that the passbands are double peaked, with the
{1700\ \AA} in particular having two almost equal components centred at $402\pm259$ and {$208\pm 174$\ km}. 
The helioseismic signal in the {1600\ \AA} band, therefore, is  in general
somewhat contaminated by flare activity, but there were no strong 
flares during the day analysed. It seems unlikely that the strong
global $p$-mode signal found by \citet{2011JPhCS.271a2058H} in both {1600\ \AA} and {1700\ \AA} bands could originate
as high as the transition region, and we therefore suspect that the
helioseismic response in this bandpass is associated with the continuum
rather than the C~{\sc iv} line and pertains to a height range not very far 
above that involved in the {1700\ \AA} band.
We should note that the calibration and registration of the AIA data are 
not optimized for helioseismology in the same way as the HMI data, so there is a possibility of position errors and small drifts, of the order of a fraction of a pixel over several days, which could affect comparisons at the pixel level but should not be a serious problem for the analysis described here, which involves smoothing over several detector pixels.

The data were obtained from the JSOC web
site, \href{http://jsoc.stanford.edu/}{http://jsoc.stanford.edu/}. The website allows the extraction of a ``tracked hg\_patch'' from each image; these patches can be chosen to be centred on a given heliographic location but 
are not remapped. For each observable we obtained a 24-hour sequence of
such patches centred on the location of AR 11072, covering the whole of 
23 May 2010. 
These patches were based on images from the {\sf aia.lev1} series and 
the (level 1.5) series {\sf hmi.V\_45s}, {\sf hmi.Ic\_45s}, and {\sf hmi.Ld\_45s}; these are the
most final calibrations of the data available at the time of writing.
The data were then remapped to a common $201\times 201$ pixel grid, 15 heliographic
degrees on each side,
evenly spaced in heliographic latitude and longitude, and interpolated
to a uniform 45-second cadence. A running mean over 15 minutes was subtracted from each remapped pixel time series to remove effects such as daily variations due to the orbit of SDO; this will also reduce the reliability of the results at the low-frequency end of the spectrum.

\subsection{Phase and Coherence Functions}
\label{sec:2.2}

The relationship between two time series in the Fourier domain can be expressed in terms of the cross-spectrum, the phase difference, and the cross-coherence, which we define as follows.

Let $P_A(\nu)$ and $P_B(\nu)$ be the complex Fourier spectra of two time series.
The ``cross spectrum'' is defined as
\begin{equation}
\label{eqcross}
{\rm CROSS}(\nu) \equiv P_{AB} =P_A(\nu)\!\stackrel{\scriptscriptstyle\bullet}{{}}\! P_B^\ast(\nu) 
\end{equation}
The phase difference $[\delta\phi_{A,B}]$ can be written as
\begin{equation}
\label{eq:eqphi}
\delta\phi_{AB}(\nu)=\arg\langle P_{AB}(\nu)\rangle.
\end{equation}
We adopt the convention that a positive value of $\delta\phi_{AB}$
means that $A$ leads $B$.
The coherence spectrum, a measure of the correlation
of the two spectra as a function of frequency, can be written as 
\begin{equation}
\label{eq:eqcoh}
{\rm{COH}_{AB}(\nu)} = {{|\langle P_{AB}(\nu)\rangle|}\over
{\sqrt{\langle |P_A(\nu)|^2 \rangle \langle |P_B(\nu)|^2 \rangle}}}.
\end{equation}
The angle brackets formally denote the expectation value; in 
practice this can be approximated by a mean over many short spectra as in the work of \citet{1994MNRAS.269..529E}, a smoothing by a running mean in the frequency domain, as used by \citet{1999ApJ...525.1042J}, or, for resolved-Sun observations \citep{2001ApJ...561..444S,2004ApJ...602..516B} an average over all azimuthal orders [$m$] for a given degree [$\ell$]. 
In the present work we use 
three-dimensional boxcar smoothing of 
the data cube of one-dimensional power spectra, equivalent to taking a running mean over
three remapped pixels in each spatial direction in turn and then over {0.3\ mHz} in temporal frequency.

\section{Images}
\label{sec:3}
In Figure~\ref{fig:1} we show the mean image for each of the observables for 23 May 2010.  
On this date the active region was somewhat mature, with a well-defined leading spot and a following-polarity region consisting mostly of strong plage with a couple of small sunspots, and it was close to the central meridian. In the Doppler-velocity panel, the supergranulation is
clearly visible; the active region disrupts the supergranular pattern and 
produces the familiar 
appearance of an upflow on one side of the penumbra and a downflow on the other,  an artefact of the projection of the horizontal Evershed flow on the line of sight. The sunspot umbra and penumbra appear as darkened areas in all of the intensity observables, but the weaker fields in the following--polarity region and surrounding network are bright in the AIA bands and in HMI $I_L$. The network fields clearly correspond to the supergranule boundaries.
For additional context, we show in Figure~\ref{fig:1a} the 
mean magnetic-field inclination from Milne--Eddington inversions of HMI magnetograms and the
AIA images at 304 and {171\ \AA}, which show the structure of the overlying coronal loops that connect the 
leading and following regions.

\section{Power Maps}
\label{sec:4}
In order to study the local power distribution in each observable, we calculated the power spectrum in each pixel of the remapped data cubes. The spectra were then smoothed over 25 bins (about 0.3~mHz) in temporal frequency and three bins in each spatial direction, and the power at each frequency was normalized by the mean power for that frequency across the whole data cube. 

We now examine the power maps for each observable in turn, taking them in 
ascending order of height of formation.

\subsection{HMI $I_C$}
Figure~\ref{fig:2} shows 
four sample
relative power maps at 3, 5, 7, and {9\, mHz}, 
and a slice through the stack of relative-power maps
at constant latitude passing through the body of the active region,
for the HMI $I_C$ data cube. The acoustic cut-off frequency in quiet Sun
is around {5\ mHz}.
In this case, and for all the other intensity observables, we divided each time series by the mean intensity for the pixel to remove the unwanted effects of brighter regions.
Power is suppressed in the active region at all frequencies; there is very little variation 
in the pattern of power suppression with frequency, although there is slightly stronger suppression in the five-minute band and above than in the granulation ($\leq 2.5\,{\rm mHz}$) range. There is no sign of any haloes of excess power.
The narrow spike of excess power within the sunspot around {5\ mHz} is seen in 
only one of the other four days examined and may be a noise artifact.
The power map shows somewhat finer detail at higher frequencies, presumably due to the way that the low-frequency limit of the $p$-mode spectrum at the $f$-mode ridge shifts towards higher spatial frequencies 
with increasing temporal frequency.

\subsection{HMI Velocity}
In Figure~\ref{fig:3} we show the results for the HMI~$V$ power. Here we see the well-known suppression of the acoustic power in the  active region in the five-minute frequency band, and a weak, broad halo of excess power surrounding the active region at {5\ mHz}. In the {7\ mHz} band the excess power has shrunk to a narrow belt outlining the active region and surrounded by a wider region of suppressed power, with a hint of a more diffuse enhancement beyond that. At this frequency, the suppression of power within the 
sunspot is much weaker and even cancelled out by the enhancement 
around the edges. Examination of the
longitude--frequency slice suggests that the halo of enhanced power contracts towards the active region with increasing frequency.

\subsection{HMI $I_L$}
Figure~\ref{fig:4} shows slices through the power maps for HMI $I_L$. 
Here, there is a strong halo of excess power -- so strong that we have used a
wider greyscale range than for the $I_C$ and V maps -- above about 6~mHz surrounding the active region, but farther out than that seen in the velocity observations. This region is beyond that where there is any detectable concentration of surface magnetic field, and has no obvious correspondence to the morphology of the overlying magnetic loops as revealed in the AIA {304\ \AA} and {171\ \AA} images. At {7\ mHz}, it occupies about the same area as the outer band of reduced power in the velocity power map. Within the halo, there are ``tendrils'' of power suppression extending out from the active region that correspond to concentrations of plage field, giving it a patchy appearance.
As in the velocity observations, this halo appears to contract with increasing frequency, and at {9\ mHz} it has collapsed to a narrow band around the sunspots,
much the same as that seen in the HMI~$V$ at {7\ mHz}.
The existence of the halo is consistent with the findings of \citet{2002A&A...387.1092J}, who saw a similar phenomenon in the MDI line-depth observations.

\subsection{AIA 1700 and {1600\ \AA}}
Finally, we show the power-map slices for AIA {1700\ \AA} (Figure~\ref{fig:5}) and
AIA {1600\ \AA} (Figure~\ref{fig:6}). As in HMI $I_L$, the power in the active region is more strongly suppressed at {5\ mHz} than in the five-minute band, and suppression is also 
visible for weak network-field concentrations in the quiet Sun.
Also, as seen in HMI $I_L$, there is an outlying halo of high-frequency excess power surrounding the active region. This appears to contradict the findings of
\citet{2003A&A...401..685M}, who did not find any excess power in these bands in TRACE observations. However, we reiterate that this power excess is at a greater distance from the active region than the halo in velocity, and also occurs
at higher temporal frequencies. Haloes of excess power have also been seen in 
ground-based observations using chromospheric lines, for example by \citet{2007A&A...471..961M} in the Na~D line and  by \citet{1992ApJ...392..739B} in Ca~K. 
The halo region contracts and weakens with increasing frequency, becoming indistinct by about {10\ mHz}, but
the collapse and encroachment on the active region seen in the HMI~$V$ and HMI $I_L$ cases do not happen
up to the {10.4\ mHz} Nyquist frequency of the {1700\ \AA} data set, 
and a re-analysis of the {1600\ \AA} data with the
full 24-second temporal resolution did not reveal any such effect at higher frequencies.

\section{Two-Dimensional Phase and Coherence maps}
\label{sec:5}

\begin{figure}
\includegraphics[width=11.0cm]{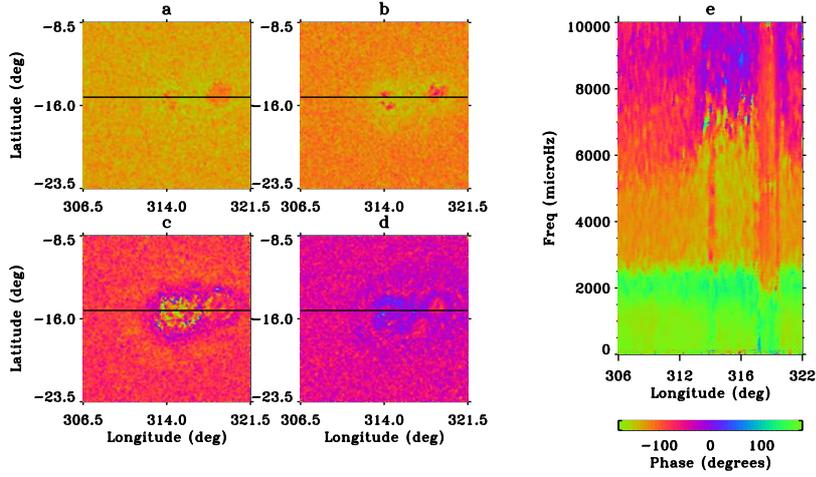}
\caption{Slices through the HMI $I_C$/HMI~$V$ phase, 
at {3\ mHz} (a), {5\ mHz} (b), {7\, mHz} (c), {9\ mHz} (d), and in longitude and temporal frequency along the
horizontal line shown in panels a\,--\,d (e). The colour-scale bar applies to all of the panels.}
\label{fig:7}
\end{figure}
\begin{figure}
\includegraphics[width=11.0cm]{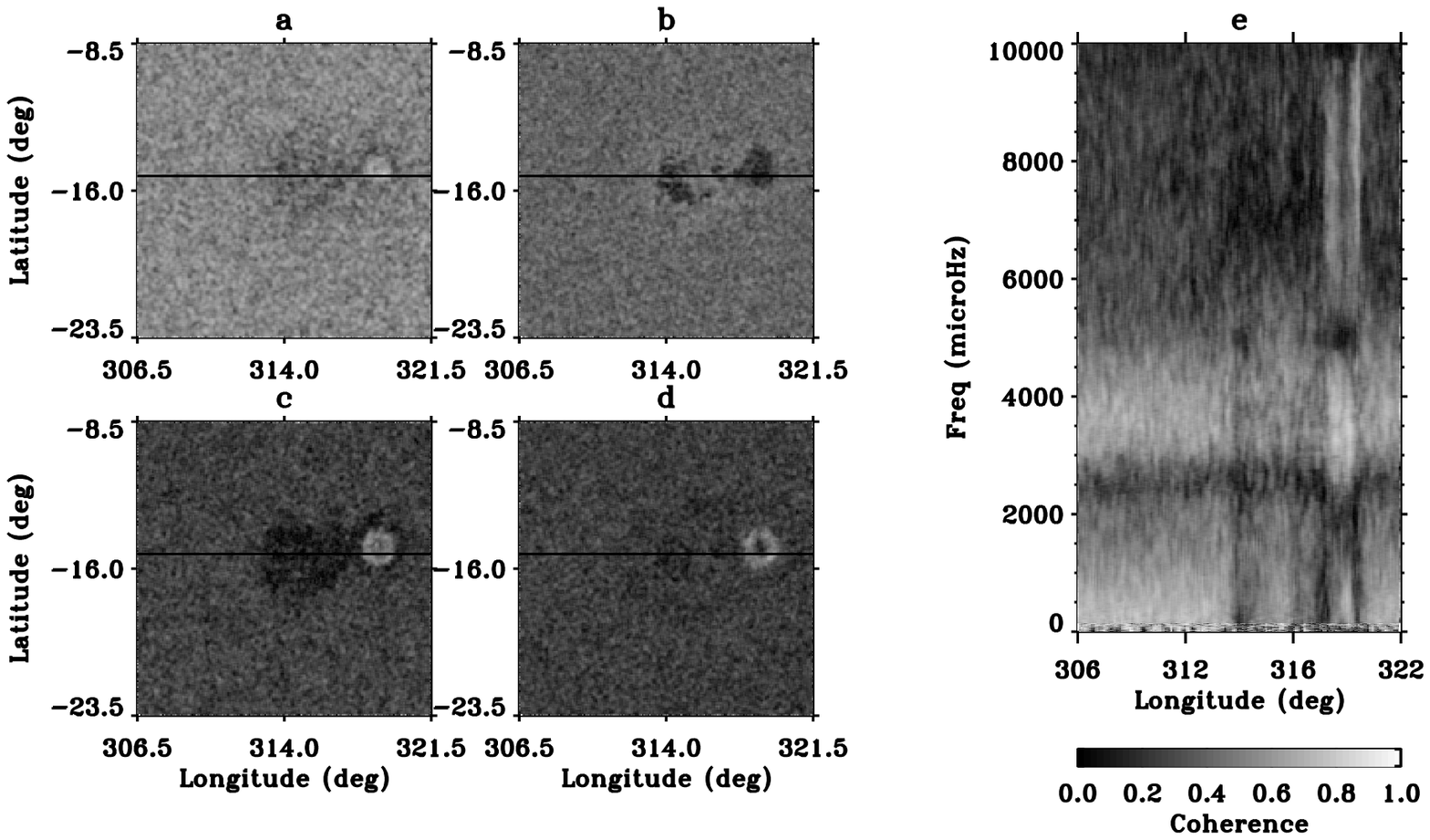}
\caption{Slices through the HMI $I_C$/HMI~$V$ coherence, 
at {3\ mHz} (a), {5\ mHz} (b), {7\ mHz} (c), {9\ mHz} (d), and in longitude and temporal frequency along the
horizontal line shown in panels a\,--\,d (e). The grey-scale bar applies to all of the panels.}
\label{fig:7a}
\end{figure}

\begin{figure}
\includegraphics[width=11.0cm]{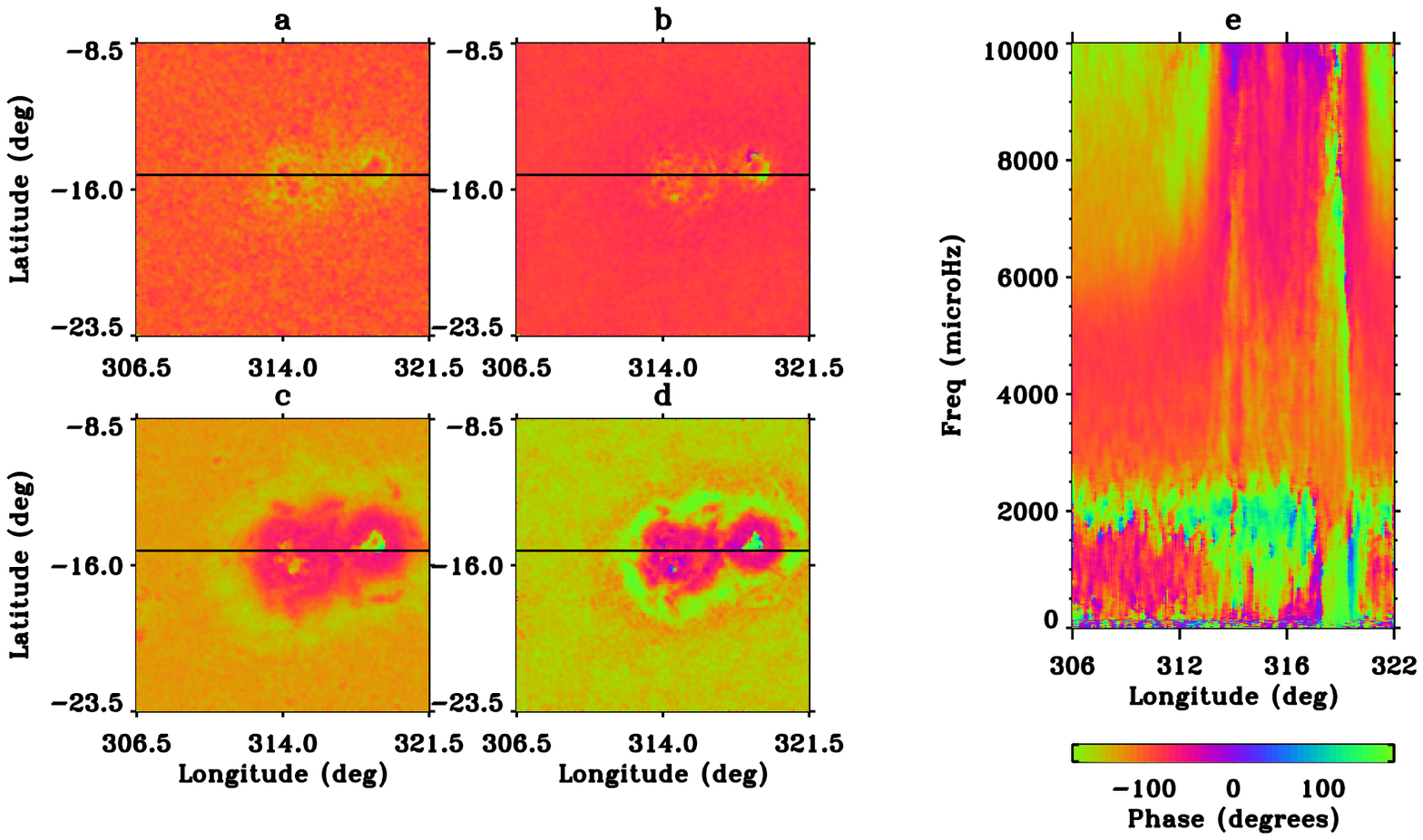}
\caption{Slices through the HMI $I_L$/HMI~$V$ phase, at {3\ mHz} (a), {5\ mHz} (b), {7\ mHz} (c), {9\ mHz} (d), and in longitude and temporal frequency along the
horizontal line shown in panels a\,--\,d (e). The colour-scale bar applies to all of the panels.}
\label{fig:8}
\end{figure}
\begin{figure}
\includegraphics[width=11.0cm]{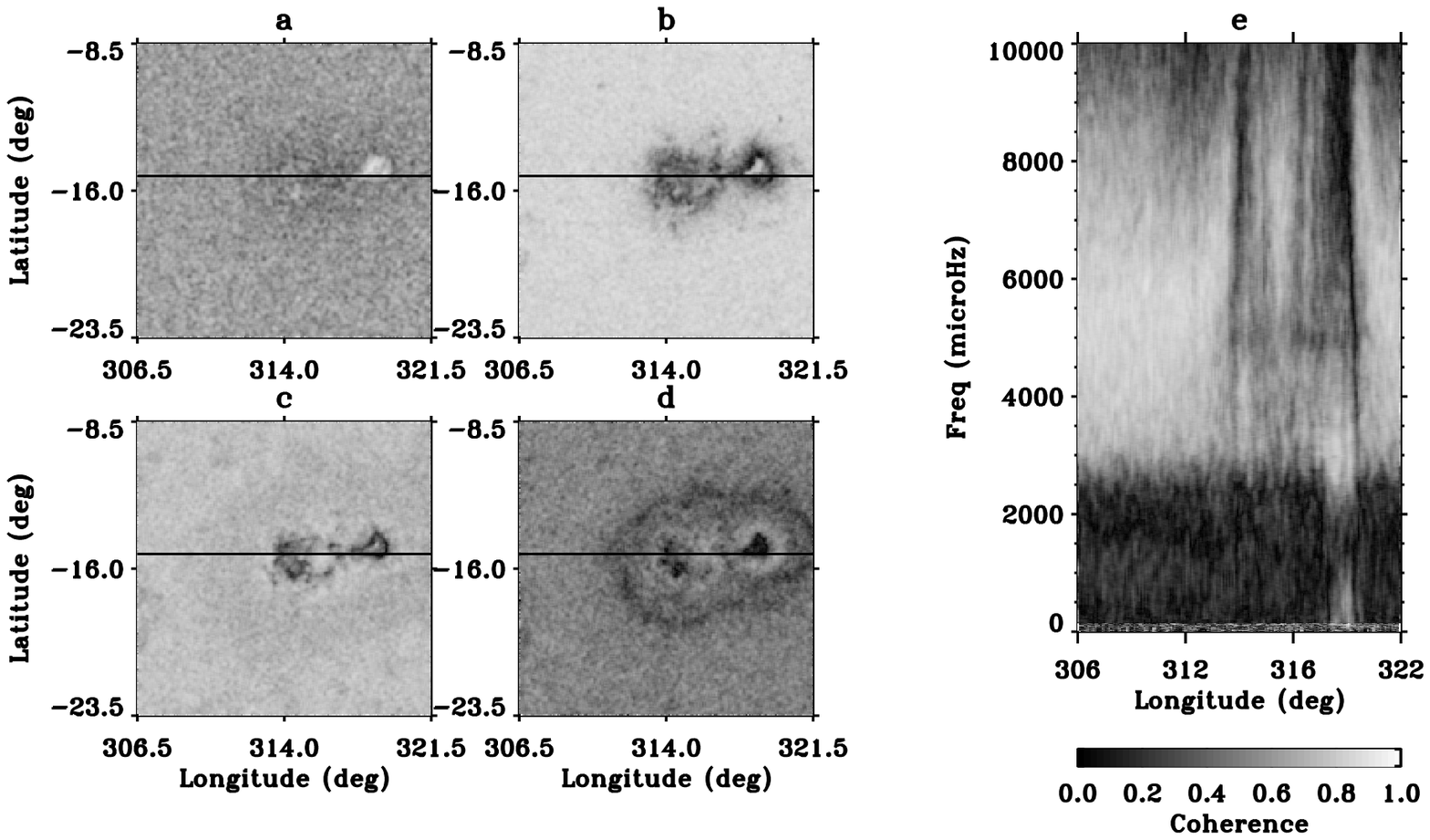}
\caption{Slices through the HMI $I_L$/HMI~$V$ coherence at {3\ mHz} (a), {5\ mHz} (b), {7\ mHz} (c), {9\ mHz} (d), and in longitude and temporal frequency along the
horizontal line shown in panels a\,--\,d (e). The grey-scale bar applies to all of the panels.}
\label{fig:8a}
\end{figure}

\begin{figure}
\includegraphics[width=11.0cm]{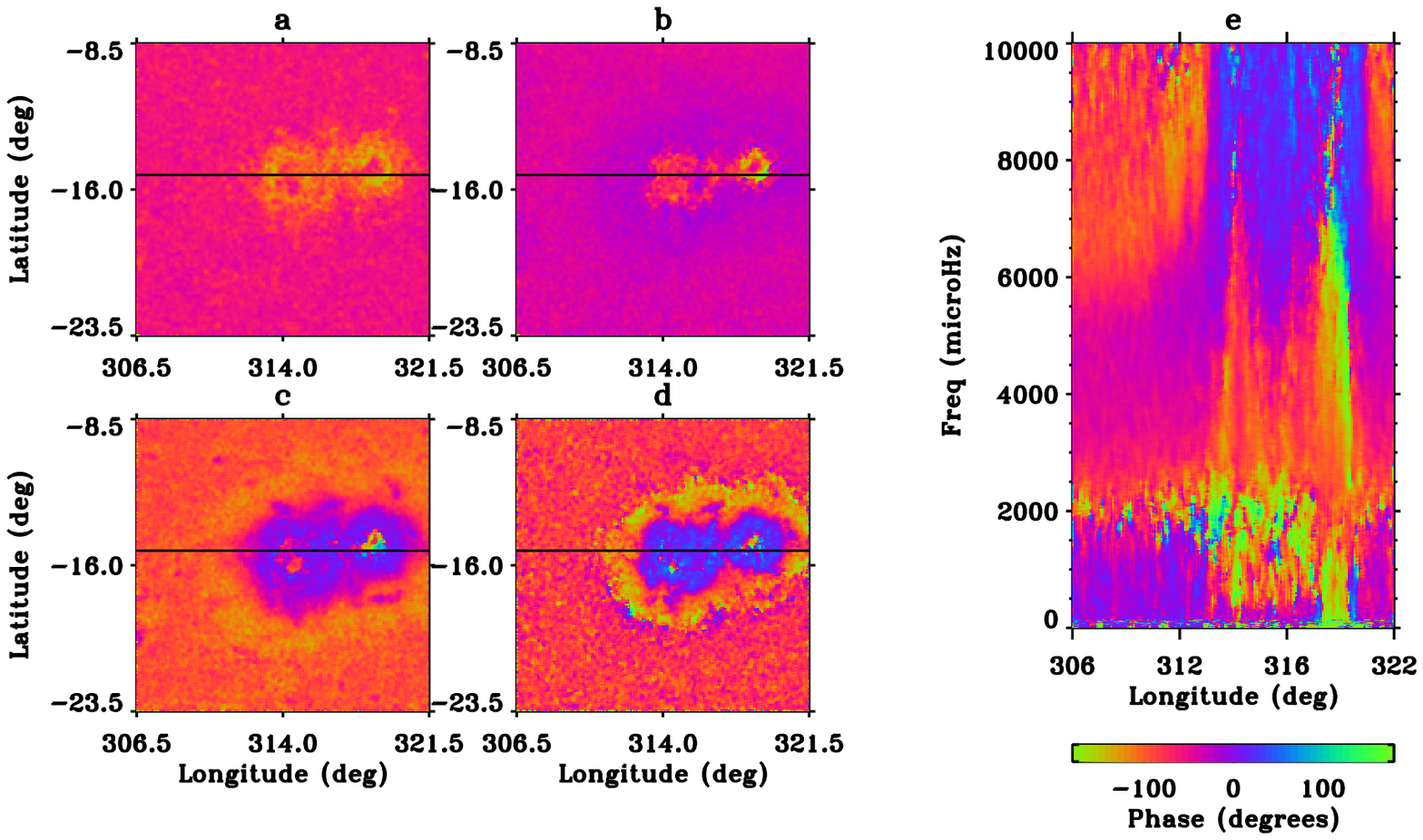}
\caption{Slices through the AIA {1700\ \AA}/HMI~$V$ phase at {3\ mHz} (a), {5\ mHz} (b), {7\ mHz} (c), {9\ mHz} (d), and in longitude and temporal frequency along the
horizontal line shown in panels a\,--\,d (e). The colour-scale bar applies to all of the panels.}
\label{fig:9}
\end{figure}
\begin{figure}
\includegraphics[width=11.0cm]{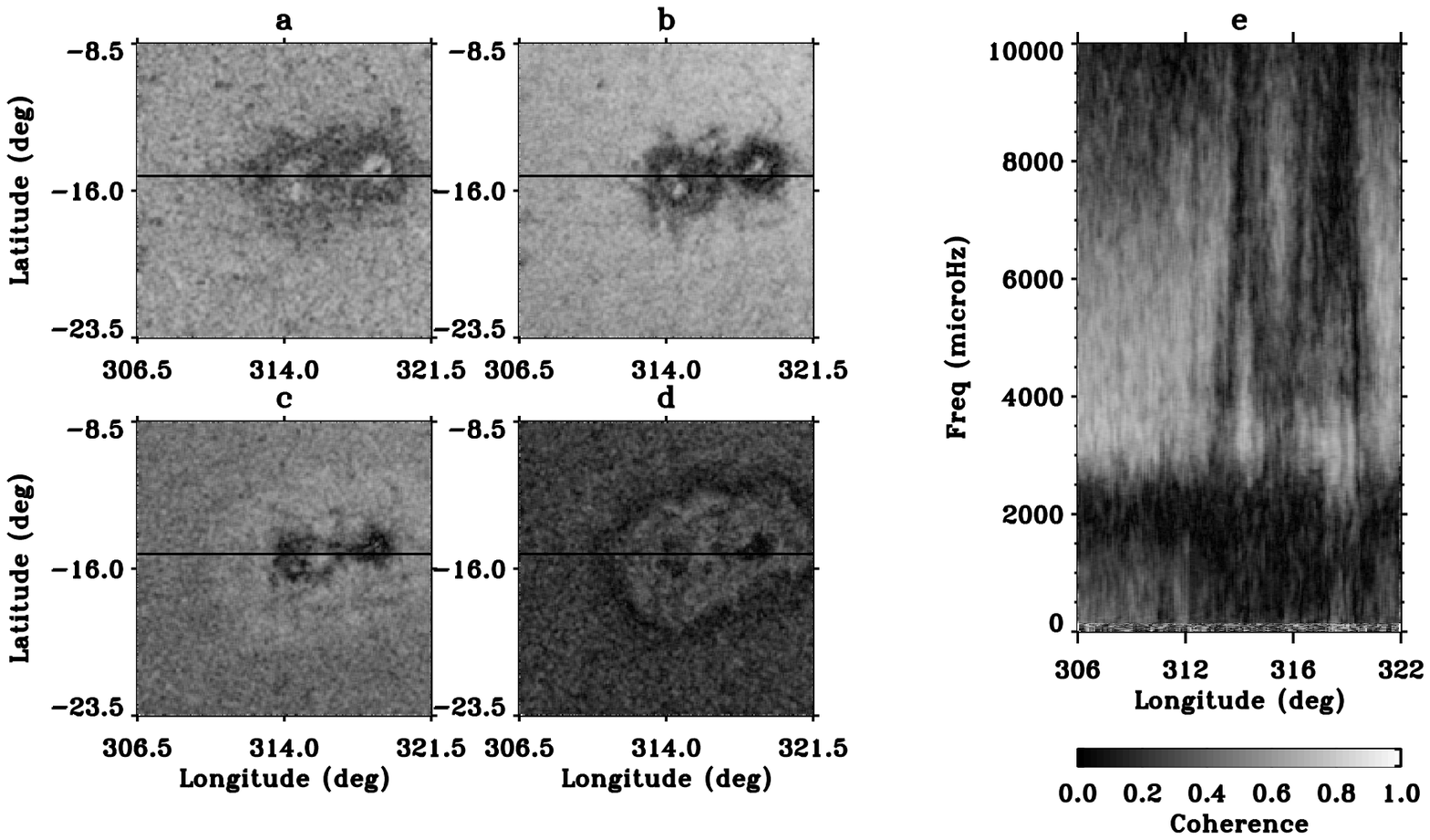}
\caption{Slices through the AIA {1700\ \AA}/HMI~$V$ coherence at {3\ mHz} (a), {5\ mHz} (b), {7\ mHz} (c), {9\ mHz} (d), and in longitude and temporal frequency along the
horizontal line shown in panels a\,--\,d (e). The grey-scale bar applies to all of the panels.}
\label{fig:9a}
\end{figure}

\begin{figure}
\includegraphics[width=11.0cm]{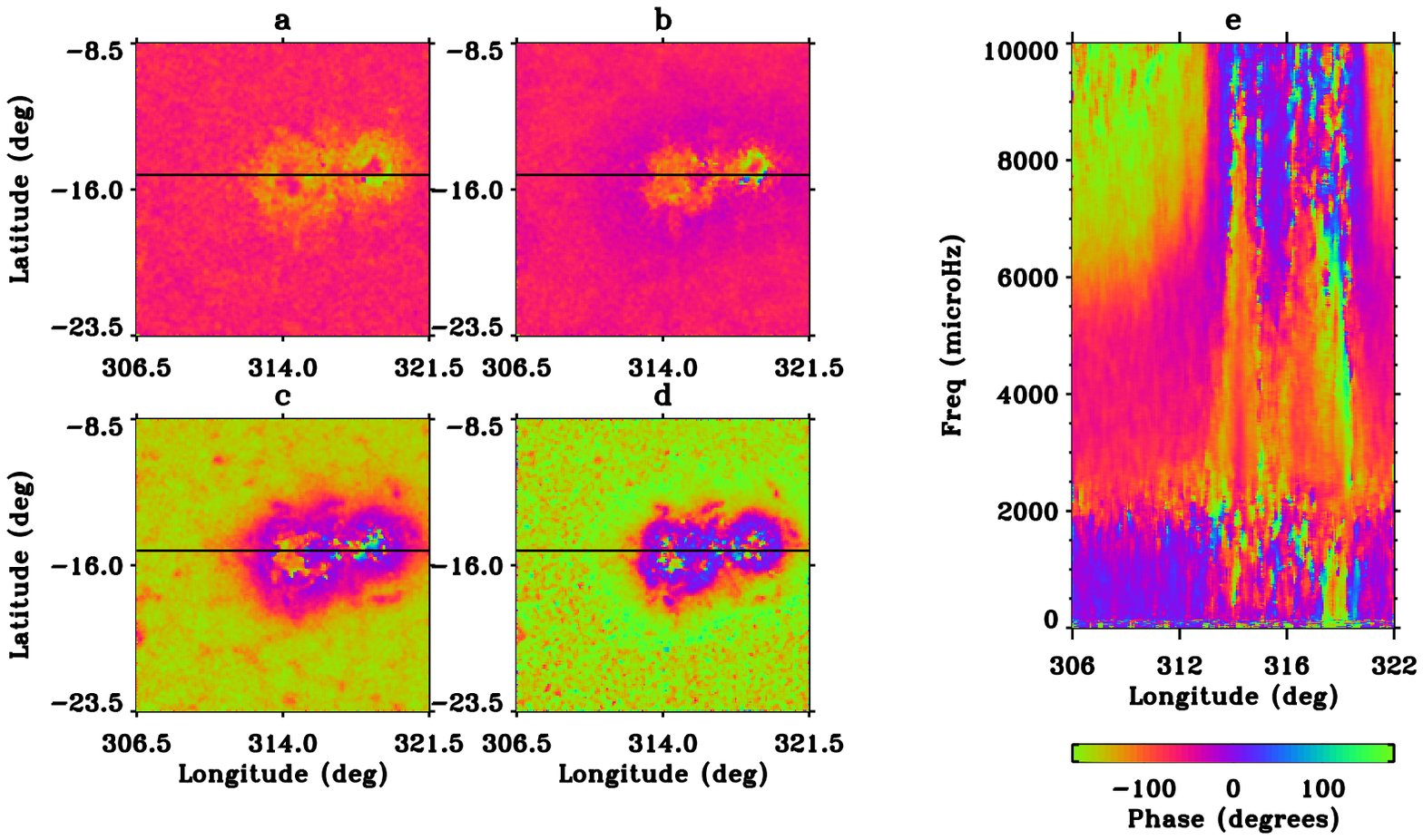}
\caption{Slices through the AIA 1600\ \AA/HMI~$V$ phase at {3\ mHz} (a), {5\ mHz} (b), {7\ mHz} (c), {9\ mHz} (d), and in longitude and temporal frequency along the
horizontal line shown in panels a\,--\,d (e). The colour-scale bar applies to all of the panels.}
\label{fig:9c}
\end{figure}
\begin{figure}
\includegraphics[width=11.0cm]{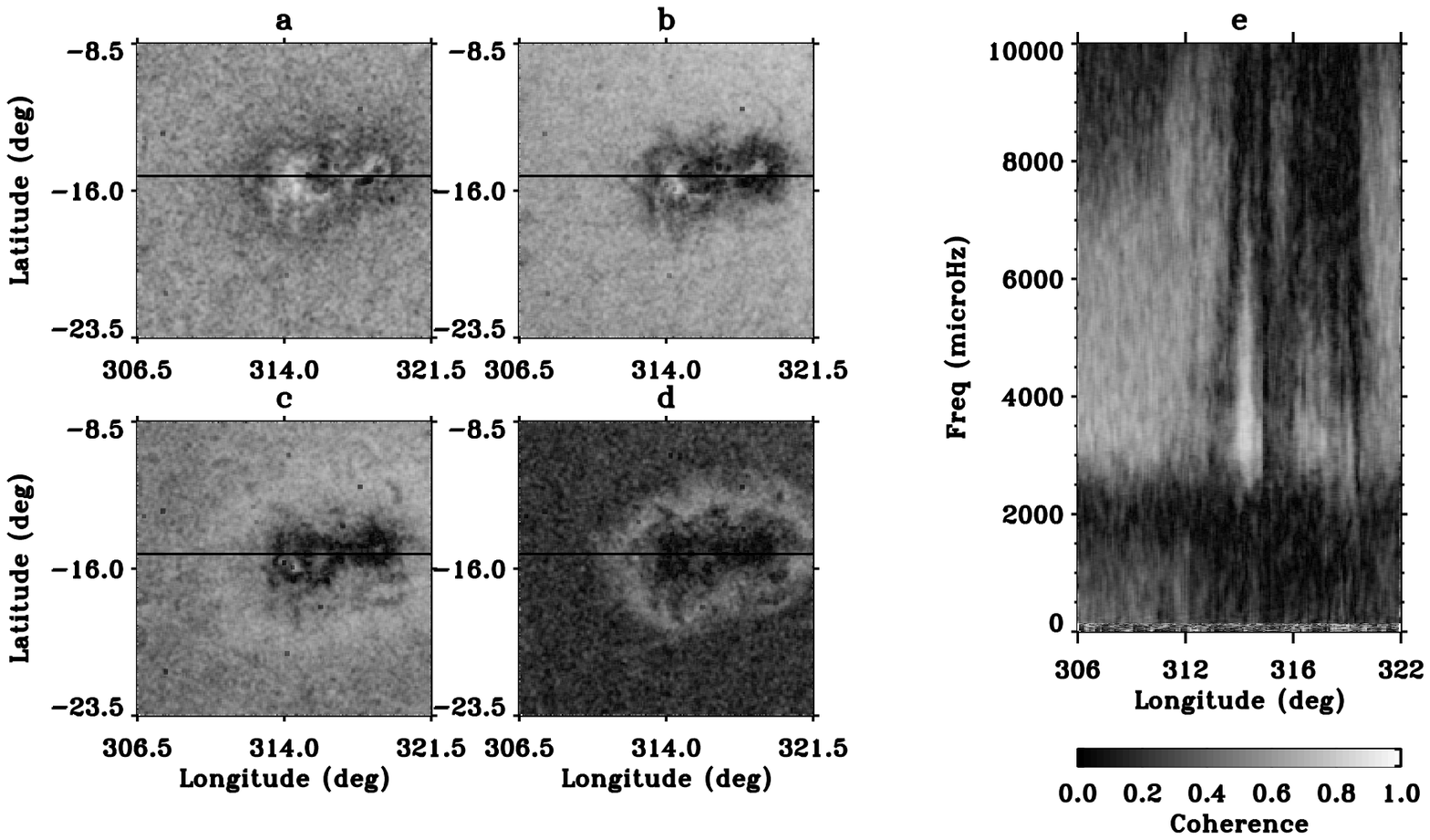}
\caption{Slices through the AIA {1600\ \AA}/HMI~$V$ coherence at {3\ mHz} (a), {5\ mHz} (b), {7\ mHz} (c), {9\ mHz} (d), and in longitude and temporal frequency along the
horizontal line shown in panels a\,--\,d (e). The grey-scale bar applies to all of the panels.}
\label{fig:9d}
\end{figure}

The phase and coherence of the individual remapped pixel time series, as a function of position and frequency, can be derived using Equations~(\ref{eq:eqphi}) and (\ref{eq:eqcoh}). When the coherence is low there is little relationship
between the two signals and we can expect the apparent phase difference 
to be close to zero after spatial averaging.

In Figures~\ref{fig:7} to \ref{fig:9d} we show these quantities
for the combination of HMI~$V$ with HMI $I_C$, HMI $I_L$, AIA {1700\ \AA}, and AIA {1600\ \AA}. The velocity
is treated as positive in the direction away from the observer.
In all cases, the coherence with the velocity is sharply reduced within the
active region, which is understandable due to the greatly reduced 
mode power and consequent reduction in signal-to-noise ratio. There is, however, a local increase in coherence in the core of the sunspot for all three variables.
This may be a real effect, but we should be cautious of common spurious, or at least complicating, effects within the sunspot umbra, such as noise or tracking errors interacting with the steep 
gradients of power and intensity, contamination by other spectral lines due to reduced temperatures, and possible
cross-talk between the HMI observables in the presence of strong fields.

The phase difference between HMI~$V$ and HMI $I_C$ shows a sharp discontinuity 
at around {2.5\ mHz}, the upper limit of the granulation signal; this phenomenon has been well known since the work of, for example, \citet{1985PhDT.......213S} as illustrated by \citet{1990A&A...236..509D}, and of \citet{1999ApJ...525.1042J}. In the core of the active region this 
boundary is shifted to lower frequencies, perhaps due to the suppression of convective motions by the surface fields. The weaker fields of the plage
area have very little effect on the $I_C$--$V$ phase in the five-minute band, but do modulate the $I_L$--$V$ phase difference.
In the HMI $I_L$ and AIA cases, the discontinuity is smaller and the low-frequency behaviour more complicated, with the phase in the plage region heavily
disrupted.

While the quiet-Sun phase difference between HMI $I_C$ and HMI~$V$ becomes slowly
less negative with increasing frequency, the opposite is true for
HMI $I_L$ and the UV bands.

Another interesting feature is that the high-frequency coherence between the HMI~$V$ and the UV intensities
is higher in the ``halo'' zone around the active region than in the
surrounding quiet Sun; this region (which, as mentioned above, does not correspond to any detectable magnetic fields) also shows a shift in the phase between these observables that 
varies smoothly with distance from the active region, going from a difference of about
$-100$ degrees for the {1700\ \AA} and $-160$ degrees for the {1600~\AA} band in quiet Sun far away from the active region at {7\ mHz} to nearly in-phase within it.
While the coherence of HMI~$V$ with AIA {1600\ \AA} and {1700\ \AA} look very similar at lower
frequencies, the 9-mHz map shows a different pattern; while for the {1700\ \AA} band the coherence is weakly enhanced over most of the extended halo region except for the active region core and an external fringe of reduced coherence, for the {1600\ \AA}  band there is a high-coherence fringe surrounding a patch of reduced coherence over the wider active region. 

On the other hand, the coherence between HMI~$V$ and HMI $I_L$ is reduced in the same region and frequency range, and the phase difference varies from $-130$ degrees in quiet Sun to $-90$ degrees in the active region. At {3\ mHz}, conversely, the phase difference between HMI~$V$ and both HMI $I_L$ and AIA {1600\ \AA} and {1700\ \AA} is more negative within the active region than in quiet Sun. It is curious that the
quiet-Sun phase is more negative than that for the AIA bands, as this would seem to imply (if we were dealing with upward-traveling waves) that the 
HMI~$I_L$ is actually probing a higher level of the atmosphere than the UV bands.

\section{Discussion}
\label{sec:6}
We have examined the spatio--temporal power distribution around an active region in a number of
HMI and AIA observables, and the phase and coherence relationships between the
intensity observables and the HMI Doppler velocity.  

Five-minute power in all observables is suppressed in the sunspot (which is dark at all wavelengths) and also in plage areas (bright in AIA bands and HMI $I_L$).
Above the acoustic cut-off frequency the behaviour is more complicated.
Power in HMI $I_C$ is suppressed in the presence of surface fields at all frequencies, while 
power in HMI~$V$ is enhanced in a narrow zone around field concentrations (especially plage) and suppressed in a wider surrounding area; these results
are consistent with earlier work using MDI.
Power in HMI $I_L$ and the  AIA bands is suppressed in areas of surface field but -- in contrast to the results of \citet{2003A&A...401..685M} -- enhanced in an extended area around the active region. In the HMI $I_L$ case, the
pattern of narrow halo, encroaching on the region where the five-minute power is suppressed, and surrounding region of suppressed power, seen for the
HMI~$V$ at {7\ mHz}, is seen instead at {9\ mHz}. For the UV bands, however, this pattern does not appear at any frequency; instead, the halo power fades with increasing frequency and vanishes above about {10\ mHz} while the power suppression remains. 

In all cases but HMI $I_C$, the regions of enhancement and suppression 
of power appear to move inwards towards the active region at increasing frequency.

The relative phase of the observables is altered around active regions. While the apparently in-phase behaviour of most observables in strong-field areas is associated with low coherence and thus of low significance, there are exceptions to this. In particular, in the 7~mHz frequency band there are areas close to the active region where the AIA {1700 \AA} and {1600 \AA} bands are both close in phase to the HMI velocity and coherent with it, whereas in the quiet Sun well away from the active region the phase difference is close to 90 degrees and the coherence is lower. This effect is not seen in the HMI $I_L$, which is otherwise very similar in its behaviour to the UV bands, although there is a phase difference between the two observables. At {9\ mHz}, the 1600 and {1700\ \AA} bands also show qualitatively different behaviour, suggesting that though there may be substantial overlap between the two wavelength bands, the highest-frequency waves are sensing different layers of the atmosphere.

Clearly, within the ``halo'' zone surrounding the active region the 
propagation and reflection of the high-frequency waves differs from that over
quiet Sun, whether due to alteration in the height of formation of the
observables, thermal changes, or the direct effect of magnetic fields in trapping, scattering or transforming the waves; also,
there are qualitative differences between the purely photospheric observables -- even at the relatively high formation level of the HMI $I_L$ -- and the UV bands. 

 In general, the higher the observable is formed, the higher the frequency at which the halo of excess power is found, and the more extensive the halo at the lowest frequency
at which it appears.
It is tempting to interpret these halo effects in terms of spreading magnetic-field canopies, as has been suggested, for example, by \citet{2007A&A...471..961M}, but we note that the 
extent of the high-frequency halo does not exactly match the morphology of the overlying field seen in the EUV images. 
The size of the halo and the inner and outer regions of suppressed power
appear to vary with frequency as well as with the height of formation 
of the observable used, contracting with increasing frequency. If the patterns of power distribution are related to spreading out of magnetic fields with height, this would imply that the higher-frequency pseudomodes are being reflected or absorbed at lower layers than the lower-frequency ones. The heights of formation themselves may also be altered in the presence of magnetic fields, further complicating the picture.

These effects no doubt also 
influence 
the changes in the local helioseismic parameters of modes above the acoustic cut-off frequency, seen for example by \citet{2004ApJ...608..562H,%
2008ASPC..383..305H}, which tend to be in the opposite sense from those experienced
by modes in the five-minute range.

Eventually, we hope to carry out multi-wavelength helioseismic analysis using 
HMI and AIA data together to extract parameters for local and global
acoustic spectra. As the results described here demonstrate, any such analysis will need to take into account the effects of active regions on the phase and coherence of the oscillations. 

The continuous, high-cadence, full-disc observations of HMI and AIA allow us to study the
behaviour of waves in the photosphere and lower chromosphere at a level of detail that has not previously been possible. These observations may help us to improve our understanding of the interaction of waves and magnetic fields in the different layers of the photosphere.  While the analysis described here does not distinguish between true waves and other short-period spatio--temporal variations, we hope in the future to examine the three-dimensional power spectrum and its phase and coherence in order to separate the waves from granulation and other background effects.

\begin{acks}
RH thanks the National Solar Observatory for computing support, and Yvonne Elsworth for useful discussions. 
We also thank A.G. Kosovichev for useful comments on the manuscript.
SDO data courtesy SDO (NASA) and the AIA and HMI consortia.
This work was partly supported by NASA grant NNH12AT11I to NSO.
This research has made use of NASA's Astrophysics Data System.
\end{acks}

\bibliographystyle{spr-mp-sola-cnd}      
\bibliography{ms_arxiv.bib}   

%
%
\end{article}
\end{document}